\documentclass[superscriptaddress,prl,twocolumn,amsmath,amssymb]{revtex4-1}
\usepackage{graphicx}
\usepackage{eurosym}
\usepackage{dcolumn}
\usepackage{bm}
\usepackage{subfigure}
\usepackage[final]{pdfpages}
\graphicspath{{./images/}}

\newcommand{\ket}[1]{\left | #1 \right \rangle}
\newcommand{\bra}[1]{\left \langle #1 \right |}

\newcommand{\braket}[2]{\mbox{$ \langle #1 | #2 \rangle $}}

\newcommand{\ignore}[1]{}

\newcommand{\be}{\begin{equation}}
\newcommand{\ee}{\end{equation}}
\newcommand{\ba}{\begin{eqnarray}}
\newcommand{\ea}{\end{eqnarray}}

\begin{document}

\title{A quantum delayed choice experiment}

\author{Alberto Peruzzo}
\affiliation{Centre for Quantum Photonics, H.H.Wills Physics Laboratory \& Department of Electrical and Electronic Engineering, University of Bristol, Bristol BS8 1UB, UK}

\author{Peter Shadbolt}
\affiliation{Centre for Quantum Photonics, H.H.Wills Physics Laboratory \& Department of Electrical and Electronic Engineering, University of Bristol, Bristol BS8 1UB, UK}

\author{Nicolas Brunner}
\affiliation{H.H. Wills Physics Laboratory, University of Bristol, Tyndall Avenue, Bristol, BS8 1TL, United Kingdom}

\author{Sandu Popescu}
\affiliation{H.H. Wills Physics Laboratory, University of Bristol, Tyndall Avenue, Bristol, BS8 1TL, United Kingdom}

\author{Jeremy L. O'Brien}
\email{Electronic email: {\tt Jeremy.OBrien@bristol.ac.uk}}
\affiliation{Centre for Quantum Photonics, H.H.Wills Physics Laboratory \& Department of Electrical and Electronic Engineering, University of Bristol, Bristol BS8 1UB, UK}

\begin{abstract}
Quantum systems exhibit particle-like or wave-like behaviour depending on the experimental apparatus they are confronted by. 
This wave-particle duality is at the heart of quantum mechanics, and is fully captured in Wheeler's famous delayed choice gedanken experiment. In this variant of the double slit experiment, the observer chooses to test either the particle or wave nature of a photon after it has passed through the slits. Here we report on a quantum delayed choice experiment, based on a quantum controlled beam-splitter, in which both particle and wave behaviours can be investigated simultaneously. The genuinely quantum nature of the photon's behaviour is tested via a Bell inequality, which here replaces the delayed choice of the observer. We observe strong Bell inequality violations, thus showing that no model in which the photon knows in advance what type of experiment it will be confronted by, hence behaving either as a particle or as wave, can account for the experimental data.
\end{abstract}

\maketitle

Quantum mechanics predicts with remarkable accuracy the result of experiments involving small objects, such as atoms and photons.
However, when looking more closely at these predictions, we are forced to admit that they defy our intuition. Indeed, quantum mechanics tells us that a single particle can be in several places at the same time, and that distant entangled particles behave as a single physical object no matter how far apart they are \cite{bell}.

In trying to grasp the basic principles of the theory, in particular to understand more intuitively the behaviour of quantum particles, some of its pioneers introduced the notion of wave-particle duality \cite{feynman}. A quantum system, for instance a photon, may behave either as a particle or a wave. However, the way in which it behaves depends on the kind of experimental apparatus with which it is measured. Hence, both aspects, particle and wave, which appear to be incompatible, are never observed simultaneously \cite{bohr}. This is the notion of complementarity in quantum mechanics \cite{scully,englert}, which is central in the standard Copenhagen interpretation, and has been intensely debated in the past. 

In an effort to reconcile quantum predictions and common sense, it was suggested that quantum particles may in fact know in advance to which experiment they will be confronted, via a hidden variable, and could thus decide which behaviour to exhibit. 
This simplistic argument was however challenged by Wheeler in his elegant `delayed choice' arrangement \cite{wheeler1,wheeler2,leggett}. 
In this gedanken experiment, sketched in Fig.~1(a), a quantum particle is sent towards a Mach-Zender interferometer. The relative phase $\varphi$ between the two arms of the interferometer can be adjusted such that the particle will emerge in output $D_0$ with certainty. That is, the interference is fully constructive in output $D_0$, and fully destructive in output $D_1$. This measurement thus clearly highlights the wave aspect of the quantum particle. However, the observer performing the experiment has the choice of modifying the above experiment, in particular by removing the second beam-splitter of the interferometer. In this case, he will perform a which-path measurement. The photon will be detected in each mode with probability one half, thus exhibiting particle-like behaviour. The main point is that the experimentalist is free to choose which experiment to perform (i.e. interference or which-path, thus testing the wave or the particle aspect), once the particle is already inside the interferometer. Thus, the particle could not have known in advance (for instance via a hidden variable) the kind of experiment it will be confronted, since this choice was simply not made when the particle entered the interferometer. 
Wheeler's experiment has been implemented experimentally using various systems, all confirming quantum predictions \cite{hellmut,lawson,kim,weihs}. In a recent experiment with single photons, a space-like separation between the choice of measurement and the moment the photon enters the interferometer was achieved \cite{jacques}.

Here we explore a conceptually different take on Wheeler's experiment. Our starting point is a recent theoretical proposal \cite{terno} of a delayed choice experiment based on a quantum-controlled beamsplitter, which can be in a superposition of present and absent. Hence, the interferometer can be simultaneously closed and open, thus testing both the wave and the particle behaviour of the photon at the same time. Here, using a reconfigurable integrated quantum photonic circuit \cite{pete}, we implement an interferometer featuring such a quantum beam-splitter, observing continuous morphing between wave and particle behaviour \cite{terno}. We point out however that this morphing behaviour can be reproduced by a simple classical model, and note that this loophole also plagues both the theoretical proposal of \cite{terno} as well as two of its recent NMR implementations \cite{roy,auccaise}.
In order to overcome this issue, we then present and experimentally demonstrate a quantum delayed choice scheme based on Bell's inequality \cite{bell64}, which allows us to test the most general classical model. 
The main conceptual novelty of this scheme is that the temporal arrangement of Wheeler's original proposal, i.e. the delayed choice of closing or not the interferometer, is not necessary anymore. Instead, we certify the quantum nature of the photon's behaviour by observing the violation of a Bell inequality. This demonstrates in a device-independent way, that is, without making assumptions about the functioning of the devices, that no local hidden variable model can reproduce the quantum predictions. In other words, no model in which the photon decided in advance which behaviour to exhibit---knowing in advance the measurement setup---can account for the observed statistics. In our experiment, we achieve strong Bell inequality violations, hence giving an experimental refutation to such hidden variable models, up to a few additional assumptions due to imperfections in our setup. 

\begin{figure}[t]
    \centering
    \includegraphics[width = 6cm]{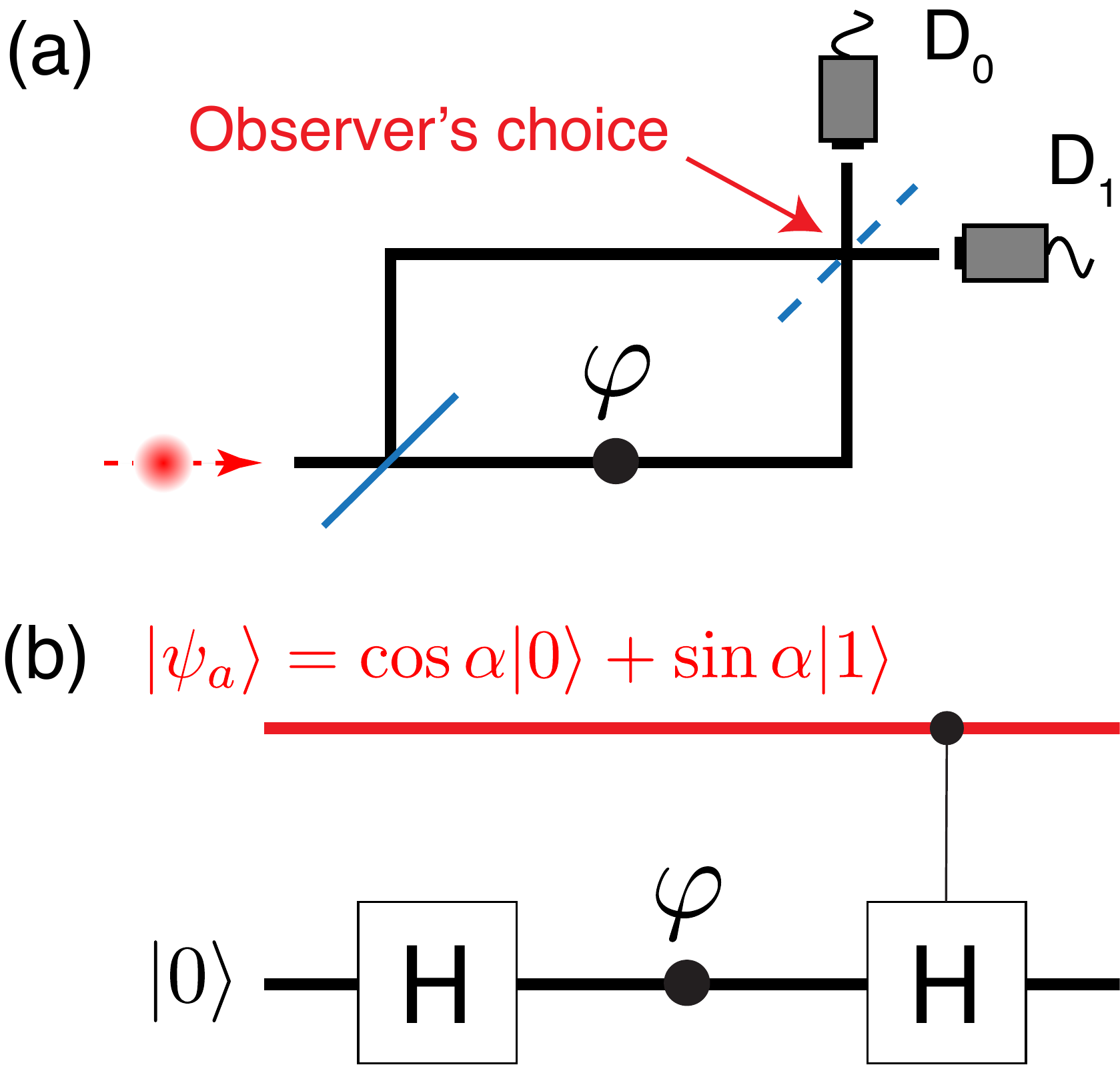}
\caption{Quantum delayed choice experiment. (a) Schematic of Wheeler's original delayed choice experiment. A photon is sent into a Mach-Zehnder interferometer. At the first beam-splitter (solid blue line), the photon is split into a superposition across both paths. Once the photon is inside the interferometer, the observer decides to close (or not) the interferometer by inserting (or not) the second beam-splitter (dashed blue line). For a closed interferometer, the statistics of the measurements at detectors $D_0$ and $D_1$ will depend on the phase $\varphi$ hence revealing the wave nature of the photon. For an open interferometer, both detectors will click with equal probability, revealing the particle nature of the photon. (b) Schematic of the quantum delayed choice experiment. The second beam-splitter is now a quantum beam-splitter (represented by a controlled-Hadamard operation), which can be set in a superposition of present and absent, by controlling the state of an ancilla photon $\ket{\psi_a}$. This allows intermediate quantum behaviour to be observed, with continuous transformation between particle and wave behaviour.  
}
\label{delayed_ch}
\end{figure}

The scheme is presented in Fig.~\ref{delayed_ch}(b). A single photon is sent through an interferometer. At the first beamsplitter, the photon evolves into a superposition of the two spatial modes, represented by two orthogonal quantum states $\ket{0}$ and $\ket{1}$. Formally, this first BS is represented by a Hadamard operation \cite{NC}, which transforms the initial photon state $\ket{0}$ into the superposition $(\ket{0}+\ket{1})/\sqrt{2}$. A phase shifter then modifies the relative phase between the two modes, resulting in the state $\ket{\psi}= (\ket{0}+e^{i\varphi}\ket{1})/\sqrt{2}$.  
Both modes are then recombined on a second BS before a final measurement in the logical (\{$\ket{0}, \ket{1}$\}) basis. In the standard delayed-choice experiment, the presence of this second BS is controlled by the observer (see Fig.~\ref{delayed_ch}(a)). 

Here, on the contrary, the presence of the second beam-splitter depends on the state of an ancillary photon. 
If the ancilla photon is prepared in the state $\ket{0}$, no BS is present, hence the interferometer is left open. Formally this corresponds to the identity operator acting on $\ket{\psi}$, hence resulting in the state 
\ba \ket{\psi_p} = \tfrac{1}{\sqrt{2}}(\ket{0}+ e^{i\varphi}\ket{1}). \ea 
The final measurement (in the $\{\ket{0},\ket{1}\}$ basis) indicates which path the photon took, in other words revealing the particle nature of the photon. The measured intensities in both output modes are equal and phase-independent, i.e $I_0=I_1=1/2$. 

If, however, the ancilla photon is prepared in the state $\ket{1}$, the BS is present and the interferometer is therefore closed. Formally this corresponds to applying the Hadamard operation to $\ket{\psi}$ resulting in the state
\ba \ket{\psi_w} = \cos\tfrac{\varphi}{2}\ket{0}- i \sin\tfrac{\varphi}{2}\ket{1}.\ea
The final measurement gives information about the phase $\varphi$ that was applied in the interferometer, but indeed not about which path the photon took. The measured intensities are $I_0 = \cos^2 \frac{\varphi}{2}$ and $I_1 = \sin^2 \frac{\varphi}{2}$.
 
\begin{figure*}[t]
    \centering
    \includegraphics[width = 16cm]{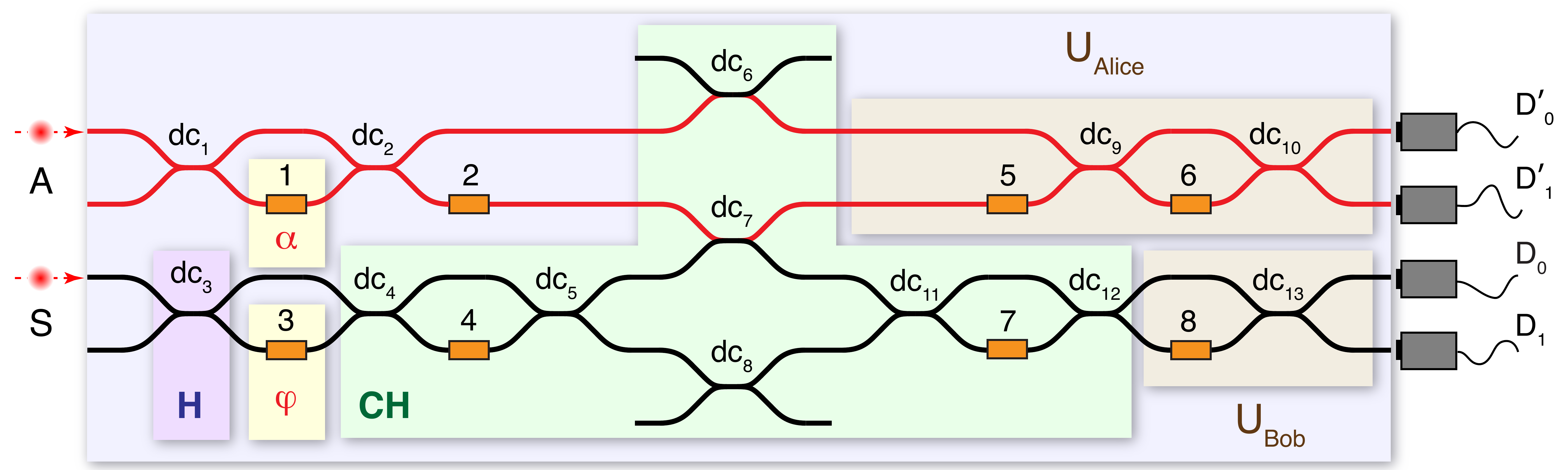}
    \vspace{-0.25cm}
\caption{Implementation of the quantum delayed choice experiment on a reconfigurable integrated photonic device.
Non-entangled photon pairs are generated using type I parametric downconversion and injected into the chip using polarization maintaining fibres (not shown). The system photon (S), in the lower part of the circuit, enters the interferometer at the Hadamard gate ($H$). A relative phase $\varphi$ is applied between the two modes of the interferometer. Then the controlled-Hadamard (CH) is implemented by a nondeterministic CZ gate with two additional MZ interferometers. The ancilla photon (A), in the upper part of the circuit, is controlled by the phase shifter $\alpha$, which determines the quantum state of the second beam-splitter, i.e a superposition of present and absent. Finally the local measurements for the Bell test are performed through single qubit rotations ($U_A$ and $U_B$) followed by APD's.
The circuit is composed of directional couplers of reflectivity 1/2 ($\text{dc}_{1-5}$ and $\text{dc}_{9-13}$), and 1/3 ($\text{dc}_{6-8}$) and resistive heaters (orange rectangles) that implement the phase shifters. See Methods section for more details.
} 
\label{scheme}
\end{figure*}

The main feature of this quantum controlled BS is that it can be put in a superposition of being present and absent. Indeed, if the ancilla photon is initially in a superposition, for instance in the state $\ket{\psi_\alpha}=\cos{\alpha}\ket{0}+\sin{\alpha}\ket{1}$, the global state of the system evolves into
\ba\label{global} \ket{\Psi_f(\alpha,\varphi)}= \cos\alpha \ket{\psi_p}_s\ket{0}_a+ \sin\alpha \ket{\psi_w}_s\ket{1}_a\ea
where the subscripts $s$ and $a$ refer to the state of the system and ancilla photons respectively.
Importantly the system and ancilla photons now become entangled, when $0<\alpha<\pi/2$.

The measured intensity at detector $D_0$ is then given by
\begin{eqnarray}
I_0(\varphi, \alpha) &=& I_p(\varphi) \cos^2\alpha + I_w(\varphi) \sin^2\alpha \nonumber \\
&=&  \frac{1}{2} \cos^2\alpha + \cos^2(\frac{\varphi}{2}) \sin^2\alpha
\label{intensity}
\end{eqnarray}
while intensity at $D_1$ is $I_1(\varphi, \alpha)=1- I_0(\varphi, \alpha)$.

We fabricated the quantum circuit shown in Fig.~2 in a silica-on-silicon photonic chip \cite{pete}. The Hadamard operation is implemented by a directional coupler of reflectivity $1/2$, equivalent to a 50/50 beam-splitter. The controlled-Hadamard (CH) is based on a non-deterministic control-phase gate \cite{ralph,hofmann}. 
The system and ancilla photon pairs are generated at 808~nm via parametric down conversion and detected with silicon APDs at the circuit's output.

\begin{figure*}[t]
    \centering
    \includegraphics[width = 16cm]{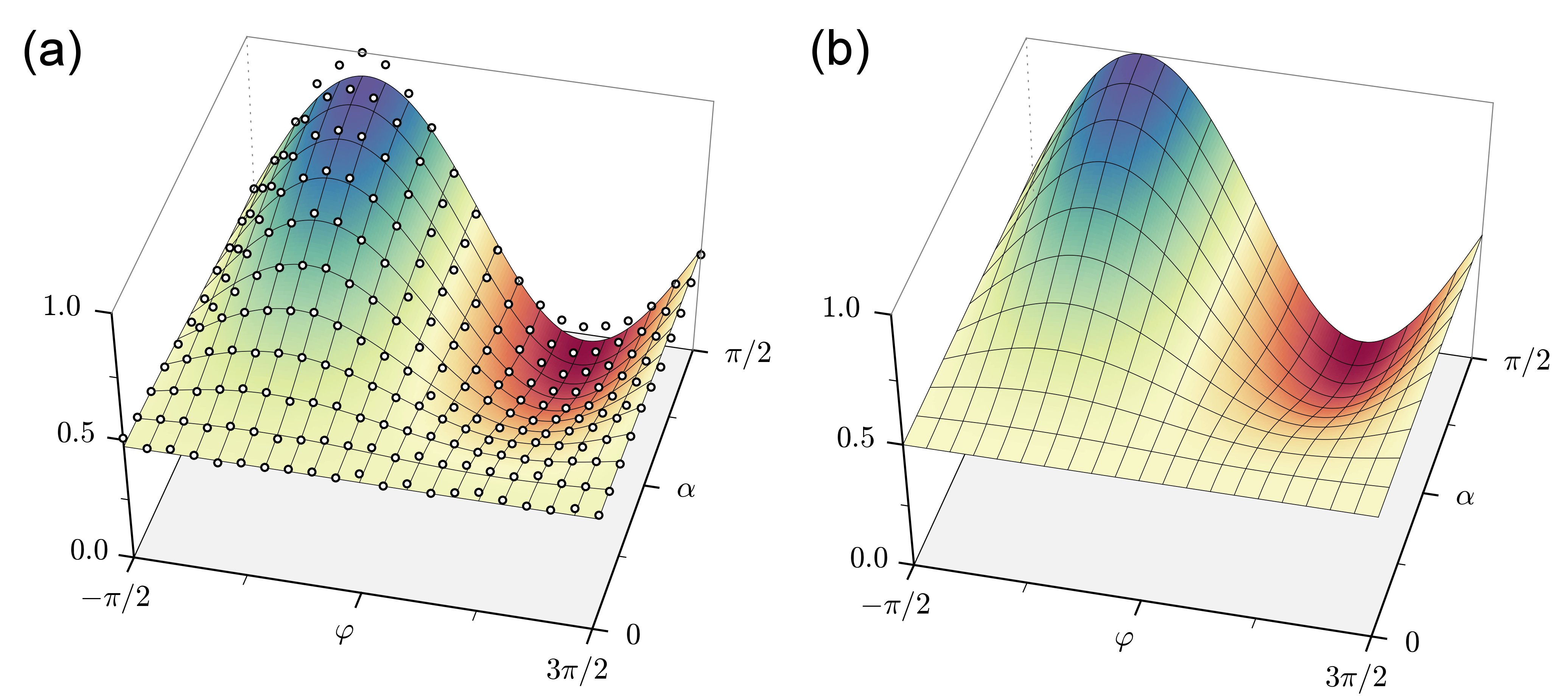}
    \vspace{-0.25cm}
\caption{Characterization of the continuous transition between wave and particle behaviour. Measured (a) and simulated (b) intensity at detector $D_0$ when continuously tuning the state of the ancilla photon $\ket{\psi_a}$. The experimental data (white dots) were fitted using equation \eqref{intensity}.  
The data shows excellent agreement with theoretical predictions.}
\label{morphing}
\end{figure*}

We first characterized the behaviour of our setup, for various quantum states of the ancilla photon. We measured the output intensities $I_{0,1}(\varphi, \alpha)$ for $\alpha \in [0,\pi/2]$, and $\varphi \in [-\pi/2, 3 \pi/2]$. In particular, by increasing the value of $\alpha$, we observe the morphing between a particle measurement ($\alpha=0$) and a wave measurement ($\alpha=\pi/2$). For $\alpha=0$, i.e. no BS, the measured intensities are independent of $\varphi$. For $\alpha=\pi/2$ the BS is present, and the data shows interference fringes.
Our results are in excellent agreement with theoretical predictions (see Fig.~\ref{morphing}).

To achieve our main goal --- that is, to refute models in which the photon knows in advance with which setup it will be confronted --- we must go one step further.
Indeed, it is important to realize that the result of Fig.~\ref{morphing} does not refute such models. The main point is that, although we have inserted the ancilla photon in a superposition, hence testing both wave and particle aspects at the same time, we have in fact not checked the quantum nature of this superposition. This is because the final measurement of the ancilla photon was made in the logical (\{$\ket{0}, \ket{1}$\}) basis. Therefore, we cannot exclude the fact that the ancilla may have been in a statistical mixture of the form $\cos^2{\alpha}\ket{0}\bra{0}+\sin^2{\alpha}\ket{1}\bra{1}$, which would lead to the same measured statistics. Hence the data can be explained by a classical model, in which the state of the ancilla represents a classical variable (a classical bit) indicating which measurement, particle or wave, will be performed. Since the state of the ancilla may have been known to the system photon in advance---indeed here no delayed choice is performed by the observer---no conclusion can be drawn from this experiment. Note that this loophole also plagues the recent theoretical proposal of Ref. \cite{terno}, as well as two of its NMR implementations \cite{roy,auccaise}.

In order to show that the measurement choice could not have been known in advance, we must ensure that our quantum controlled beam-splitter behaves in a genuine quantum way. In particular, we will ensure that it creates entanglement between the system and ancilla photons, which is the clear signature of a quantum process. The global state of the system and ancilla photons, given in equation \eqref{global}, is entangled for all values $0<\alpha <\pi/2$. Note that $\braket{\psi_p}{\psi_w} \sim \cos{\varphi}$, hence the degree of entanglement depends on $\varphi$ and $\alpha$; in particular for  
$\alpha=\pi/4$ and $\varphi=\pi/2$, the state \eqref{global} is maximally entangled.

In order to certify the presence of this entanglement, we will test the Clauser-Horne-Shimony-Holt (CHSH) Bell inequality \cite{chsh}, the violation of which will imply in a device-independent way that the measured data could not have been produced by a classical model. In the CHSH Bell scenario, each party (here Alice holds the system photon while Bob holds the ancilla photon) chooses among two possible measurement settings, denoted $x=0,1$ for Alice and $y=0,1$ for Bob. 
Each measurement is dichotomic, i.e. giving a binary result $A_x=\pm1$ and  $B_y=\pm1$. The CHSH inequality then reads
\be 
S=\langle A_0B_0\rangle + \langle A_0B_1\rangle + \langle A_1B_0\rangle - \langle A_1B_1\rangle \leq 2 
\label{bell}
\ee
This represents a Bell inequality in the sense that any local model must satisfy it.

Indeed, this inequality can be violated by making judiciously chosen local measurements on certain entangled states. We measured $S$ for the output state $\ket{\Psi_{f}(\alpha, \varphi)}$, for 
$\alpha \in [0,\pi/2]$, and $\varphi \in [-\pi/2, 3 \pi/2]$. We tailored the local measurement operators of Alice and Bob (adjusting phase shifters 5, 6 and 8, see Appendix for details) for the maximally entangled state $\ket{\Psi_f(\alpha=\pi/4,\varphi=\pi/2)}$. Hence, for this state, we expect the maximal possible violation of the CHSH inequality in quantum mechanics, namely $S=2\sqrt{2}$ \cite{tsirelson}. It is interesting to note that the choice of apparatus in Wheeler's original setup, is here, in some sense, replaced by the choice of measurement settings for the Bell test. The latter choice is nevertheless conceptually different from the former in that it can be performed after the photon left the interferometer.

\begin{figure*}[t]
    \centering
    \includegraphics[width = 16cm]{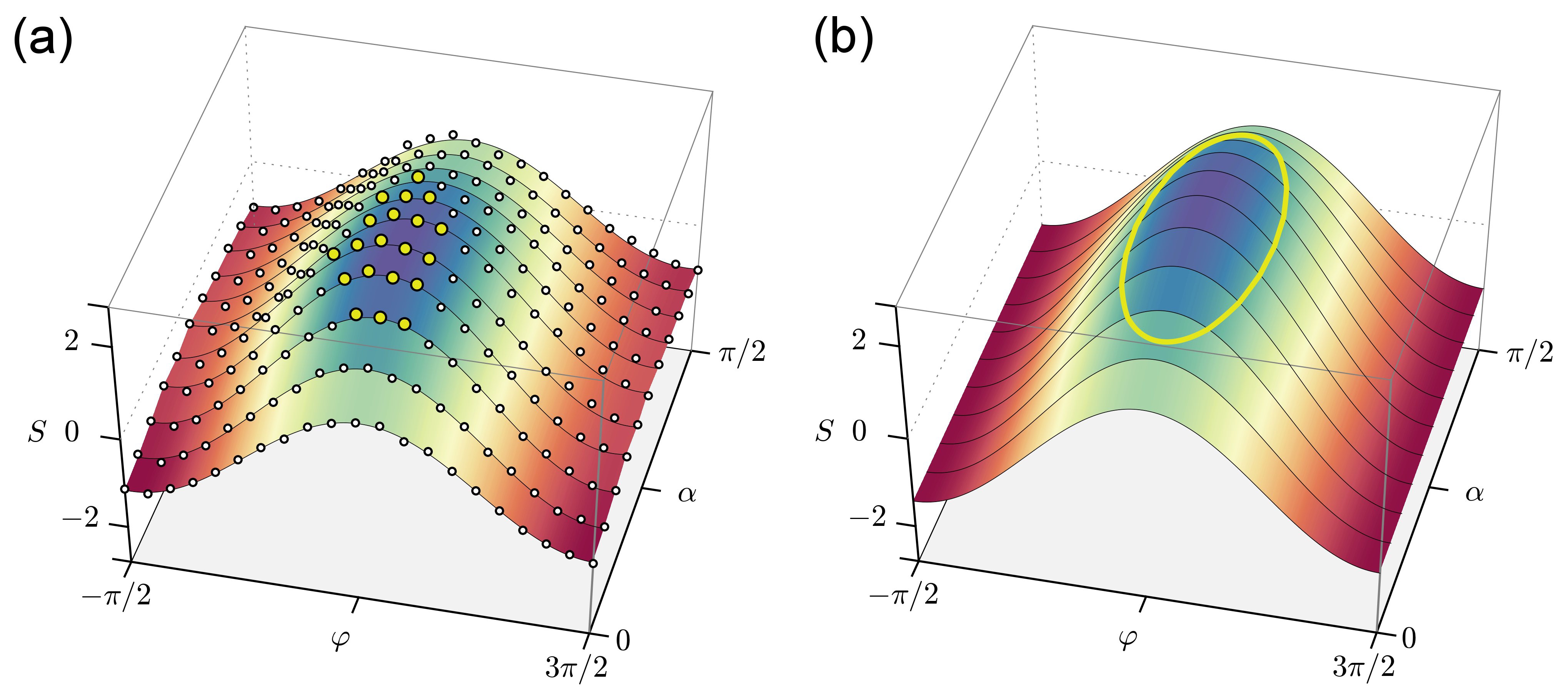} 
    \vspace{-0.25cm}
\caption{Experimental Bell-CHSH inequality test. Measured (a) and simulated (b) Bell-CHSH parameter $S$ (equation \eqref{bell}). When the CHSH inequality is violated, i.e. $S>2$ (yellow dots in (a) and yellow circle in (b)), no local hidden variable model can explain the observed data, hence demonstrating genuine quantum behaviour. The maximal experimental violation ($S=2.45\pm0.03$) is achieved for 
$\alpha=\pi/4$ and $\varphi=\pi/2$, as expected. The data is in excellent agreement with theoretical predictions.}
\label{Bell}
\end{figure*}

Experimentally we observe a maximal violation of $S=2.45\pm0.03$, for 
$\alpha=\pi/4$ and $\varphi=\pi/2$, which is in good agreement with theoretical predictions (see Fig.~\ref{Bell}). Therefore, our data could not been accounted for by any model in which the system photon would have known in advance whether to behave as a particle or as a wave. 
However, for this claim to hold without making further assumptions, a loophole-free Bell inequality violation is required. Indeed this is not the case in our experiment, as in all optical Bell tests performed so far, which forces us to make a few additional assumptions. We make the standard fair-sampling assumption (allowing us to discard inconclusive results, and post-select only coincidence events), which must here be slightly strengthened because of the nondeterministic implementation of the controlled Hadamard operation.
We must also assume independence between the photon source and the choice of measurement setting used in the Bell inequality test. As usual, if the photons could know in advance the choice of measurement setting in the Bell test, then a local model can mimic Bell inequality violations.
In the future it would be interesting to perform a more refined experiment in which these assumptions could be relaxed.

In conclusion we have reported on a quantum delayed choice experiment, giving a novel demonstration of wave-particle duality, Feynman's 'one real mystery' in quantum mechanics. In our experiment, the delayed choice of Wheeler's proposal is replaced by a quantum controlled beam-splitter followed by a Bell inequality test. In this way we demonstrate genuine quantum behaviour of single photons. 
The demonstration of a quantum controlled beam-splitter shows that a single measurement device can continuously tune 
between particle and wave measurements, hence pointing towards a more refined notion of complementarity in quantum mechanics \cite{terno,tang,qureshi}.

\emph{Acknowledgements.} We thank R.~Ionicioiu, S.~Pironio, T.~Rudolph, N.~Sangouard, and D.R. Terno for useful discussions, and acknowledge financial support from the UK EPSRC, ERC, QUANTIP, PHORBITECH, Nokia, NSQI, the Templeton Foundation, and the EU DIQIP. J.L.OÕB. acknowledges a Royal Society Wolfson Merit Award.\\

\emph{Note added.} We note a related work of Kaiser \emph{et al.} \cite{kaiser} who performed a similar quantum delayed choice experiment.

\clearpage

\section*{Appendix}

\begin{figure}[b]
    \centering
    \includegraphics[width = 8cm]{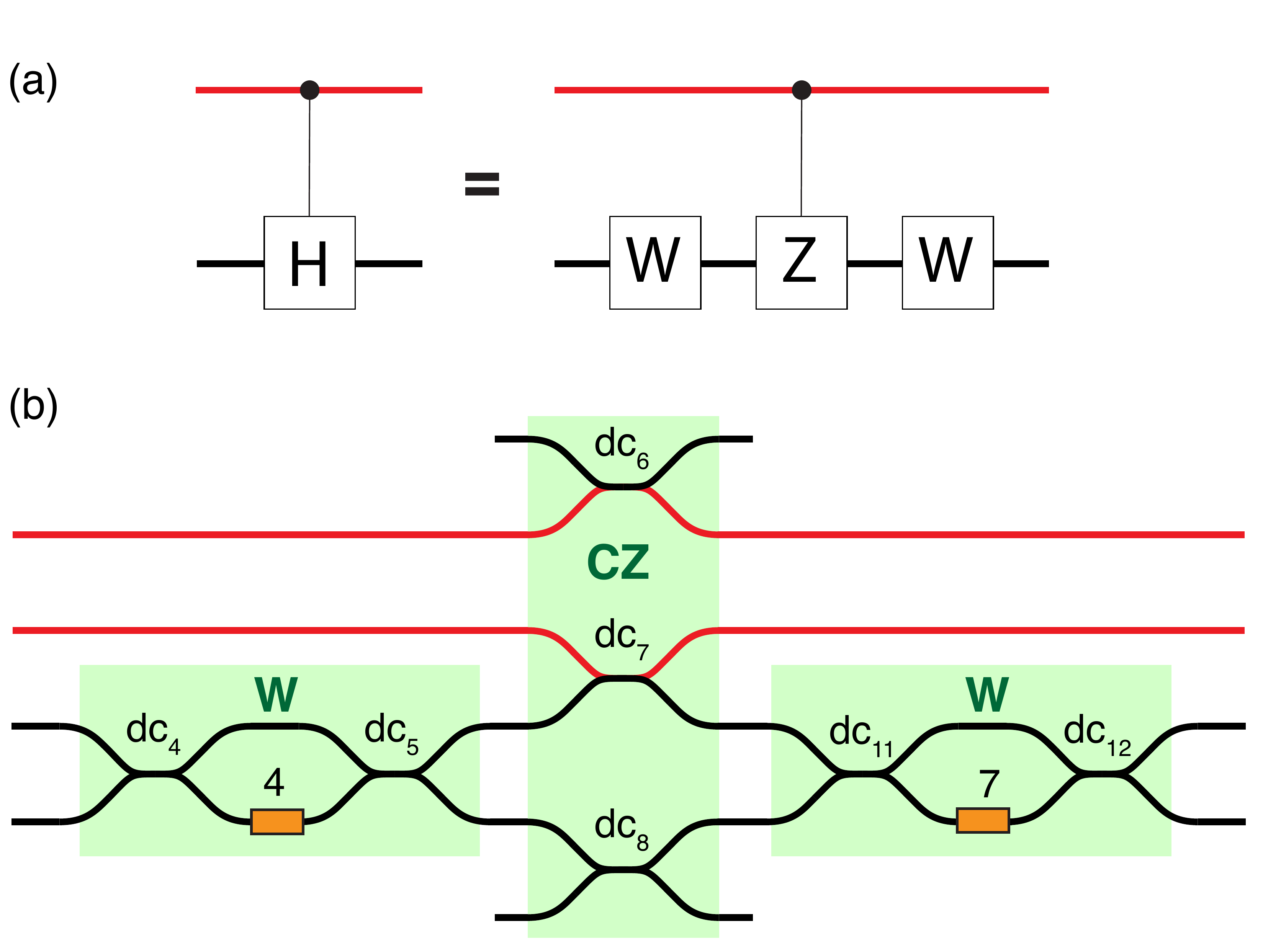}
    \vspace{-0.25cm}
\caption{Integrated photonics implementation of the CH gate. (a) Block diagram of the CH decomposition, (b) implementation using an integrated photonics device.}
\label{CH_implementation}
\end{figure}

\noindent\textbf{CH gate decomposition}\\
The CH gate is equivalent to a two-qubit CZ gate and two single-qubit W gates as shown in Fig.~\ref{CH_implementation}(a). The W gate, described by the matrix
 \begin{equation}
\mbox{W} = \left(\begin{array}{cc}\cos \frac{\pi}{8} & \sin \frac{\pi}{8} \\\sin \frac{\pi}{8} & -\cos \frac{\pi}{8}\end{array}\right),
\end{equation}
is equivalent to the operation of a beamsplitter with reflectivity $\eta = \cos^2(\frac{\pi}{8})$. This was 
 achieved, up to a non measurable global phase, by setting phase shifters 4 and 7 ---in the Mach-Zehnder interferometers indicated with W in Fig.~\ref{CH_implementation}(b)--- to $\frac{5\pi}{4}$ and $\frac{3\pi}{4}$ respectively.

\vspace{5mm}

\noindent\textbf{Measurements for the Bell test} \\
Alice and Bob's local measurement operators on the maximally entangled state \eqref{global} (with $\alpha=\pi/4$ and $\varphi=\pi/2$) were performed through single qubit rotations and single photon detection. For this state, the optimal measurement settings for the CHSH inequality are given by

\begin{equation}
A_1 = -Z\mbox{, }~ A_2 = \frac{-X-Y}{\sqrt{2}}
\end{equation}
and
\begin{equation}
B_1 = \frac{X-Z}{\sqrt{2}} \mbox{, }~ B_2 = \frac{-X-Z}{\sqrt{2}} 
\end{equation}
for Alice and Bob respectively (where $X$, $Y$, $Z$ are the usual Pauli matrices). 
In the experiment, Alice and Bob's measurements were implemented by setting ($\phi_5=-\pi/2, \phi_6=0$) for $A_1$,  ($\phi_5=\pi/4, \phi_6=\pi/2$) for $A_2$,  ($\phi_8=\pi/4$) for $B_1$,  and ($\phi_8=-\pi/4$) for $B_2$.  
\end{document}